\documentclass[10pt,a4paper]{article}
\usepackage{graphicx}
\usepackage{rotate,graphics,amssymb,amsmath}

\begin{document}
\textwidth=135mm
 \textheight=200mm

\begin{center}
{\bfseries Gravitational microlensing: results and perspectives in
brief}

\vskip 5mm
A.F.~Zakharov$^{\dag,\ddag,\S}$%

\vskip 5mm

{\small {\it $^\dag$National Astronomical Observatories of CAS, Beijing 100012, China}} \\
{\small {\it $^\ddag$Institute Theoretical and Experimental Physics,
Moscow, 117259, Russia}}
\\
{\small {\it $^\S$Joint Institute for Nuclear Research, 141980
Dubna, Russia}}
\\

\end{center}

\vskip 5mm

\centerline{\bf Abstract}
 Basics  of standard theory of
microlensing are introduced. Results of microlensing observations
toward Magellanic Clouds  and relations with dark matter (DM)
problem in our Galaxy are described. Pixel microlensing observations
and recent discoveries of planets with microlensing observations are
listed.

\vskip 10mm


Gravitational lensing is based on a simple physical phenomenon that
in a gravitational field  a light trajectories are bent  (in some
sense a gravitating body attracts photons). In the first time this
fact was discussed by I.~Newton \cite{Newton04}, but  the first
derivation of the light bending angle was published by J.~Soldner in
the framework of Newtonian gravity \cite{Soldner04}. In general
relativity (GR) using
 a weak gravitational field approximation the correct bending angle
is described by the following expression obtained by Einstein
\cite{Einstein_16} just after his formulation of GR
\begin{eqnarray}
\delta \varphi=\frac{4GM}{c^2 p},
  \label{eqs5}
\end{eqnarray}
where $M$ is a gravitating body mass, $p$ is an impact parameter,
$c$ is a speed of light, $G$ is the Newton constant. If $M=M_\odot$
and $p=R_\odot$ are Solar mass and radius respectively, the angle is
equal to $1.75''$. In 1919 the law was firstly confirmed by
A.~Eddington for observations of light ray bending by the Solar
gravitational field near its surface. Therefore, the Einstein
prediction about light bending was confirmed by observations very
soon after its discovery.

Using Eq.~(\ref{eqs5}) one can introduce the gravitational lens
equation
\begin{eqnarray}
{\vec{\eta}} ={D_s}{\vec{\xi}}/D_d  - {D}_{ds}
\vec{\Theta}(\vec{\xi}),
\end{eqnarray}
where $D_s$ is a distance between a source and observer,  $ D_d$ is
a distance between a gravitational lens and observer, $ D_{ds}$ is a
distance between a source and a lens, $\vec{\eta},\vec{\xi}$ define
coordinates in source and lens planes, respectively, and
\begin{eqnarray}
\vec{\Theta}(\vec{\xi})= 4GM \vec{\xi}/{c^2 \xi^2}. \label{eq1_4}
\end{eqnarray}
Vanishing the right hand side of Eq.~(\ref{eq1_4}), $\vec{\eta}=0$,
we obtain the so-called Einstein -- Chwolson radius
 $\xi_0=
\sqrt{{4GM}{D_dD_{ds}}/({c^2}{D_s})}$ \cite{SEF}\footnote{Chwolson
described circular images \cite{Chwolson_24} and Einstein obtained
basic expressions for gravitational lensing \cite{Einstein_36}.
Moreover, it was found that Einstein analyzed gravitational lens
phenomenon in his unpublished notes in 1912 \cite{Renn_97}.} and the
Einstein-- Chwolson angle $\theta_0=\xi_0/D_d$. If $D_s \gg D_d$, we
have
 \begin{eqnarray}
\theta_0 \approx 2'' \times
10^{-3}\left(\frac{GM}{M_{\odot}}\right)^{1/2} \left(\frac{\rm
kpc}{D_d}\right)^{-1/2}. \nonumber
\end{eqnarray}
If a gravitational lens is one of the closest galaxies with
$M=10^{12} M_{\odot}$, at a distance $D_d=100$~kpc, we have
 $\theta_0 \approx 200''$.
According to  a standard terminology if a lens mass is about
$M_{\odot}$ we call this lensing regime like microlensing one
because for cosmological distances both lenses and sources typical
angular distances between images are about microarcseconds.

We could introduce dimensionless variables
\begin{equation}
\vec{x} = \vec{\xi}/\xi_0, \quad
 \vec{y} = {D_s} \vec{\eta}/(\xi_0D_d),
\quad
\vec{\alpha} = \vec{\Theta} D_{ds}D_{d}/(D_s\xi_0),
\end{equation}
then we have gravitational lens equation in the dimensionless form:
\begin{eqnarray}
\vec{y} = \vec{x} - \vec{\alpha}(\vec{x}) \quad {\rm or} \quad
\vec{y} = \vec{x} - \vec{x}/{x^2}. \label{GL_eq1}
\end{eqnarray}
\begin{figure}[!t]
\begin{center}
\includegraphics[width=4.4cm]{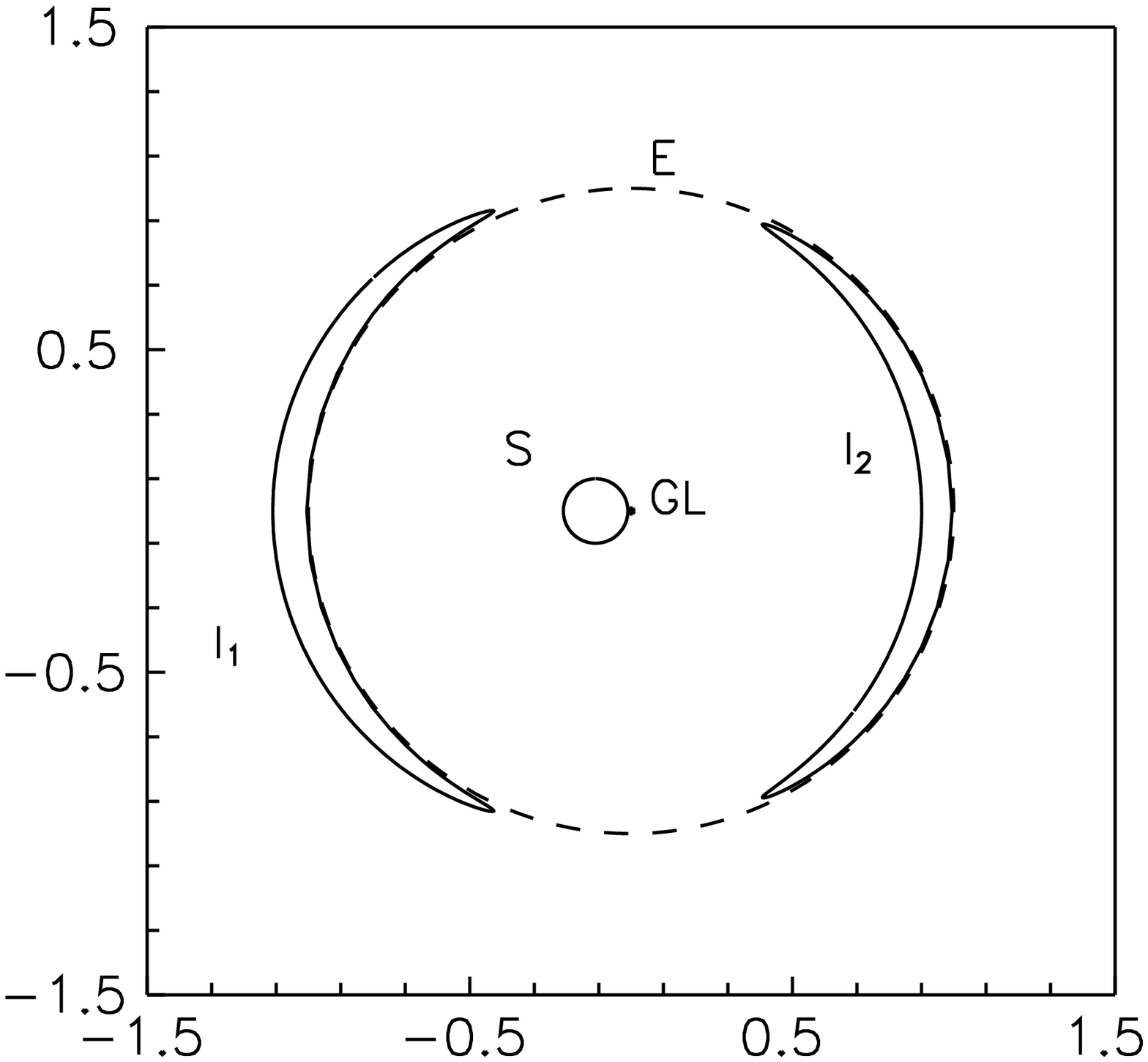}
\hspace{2.5cm}
\includegraphics[width=4.4cm]{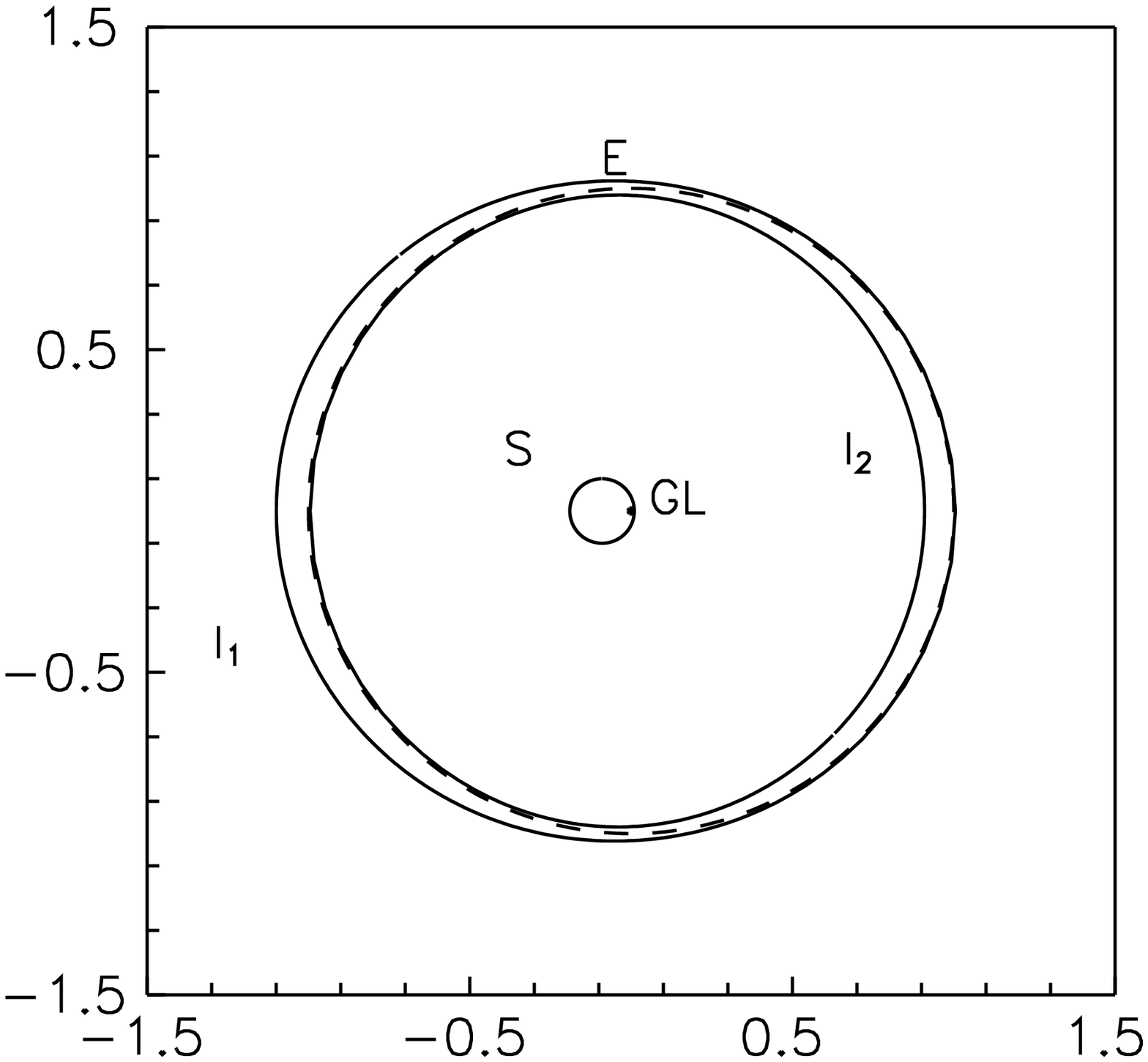}
\end{center}
\vspace{-0.4cm}
 \caption{Image formation for a circular source S with a
radius $r=0.1$ and different distances $d$ between a source center
and gravitational lens on the celestial sphere for $d=0.11$ (left
panel) and $d=0.09$ (right panel), where $I_1$ and $I_2$ are images,
E is the Einstein -- Chwolson ring, GL is a position of
gravitational lens on the celestial sphere. } \label{prot2_3}
\end{figure}
The gravitational lens effect is a formation of several images
instead of one \cite{Zakharov97}. We have two images (or one ring)
for the Schwarzschild point lens model as one can see in
Fig.~\ref{prot2_3}. The total area of the two images is larger than
a source area. The ratio of these two areas is called gravitational
lens amplification $A$. For example, if a circular source with a
radius $r$ and its area $\pi r^2$ is located near a position of
gravitational lens on a celestial sphere then an area of ring image
is equal to $2\pi r$ and therefore magnification is $2/r$ (thus one
could calculate an asymptote for $A \sim 2/r$ in a limit $r
\rightarrow 0$ by the geometrical way). That is a reason to call
gravitational lensing as gravitational focusing. As one can see the
angular distance between two images is about angular size of
so-called the Einstein -- Chwolson cone with the angle $2\theta_0$
(it corresponds to the Einstein -- Chwolson diameter). If the source
$S$ lies on the boundary of the Einstein -- Chwolson cone, then we
have  $A=1.34$. Note, that the total time of crossing the cone is
$T_0$. The microlensing time is defined typically as a half of $T_0$
$$
T_0=3.5~{\rm months} \cdot \sqrt{\frac{M}{M_{\odot}}
\frac{D_{d}}{10\, {\rm kpc}}} \cdot \frac{300\, {\rm km/s}}{v},
$$
where $v$ is the perpendicular component of a velocity of a dark
body. If we suppose that the perpendicular component of a velocity
of a dark body is equal to $\sim 300$ km/s (that is a typical
stellar velocity in Galaxy), then a typical time of crossing
Einstein cone is about 3.5 months. Thus, a luminosity of a source
$S$ is changed with the time. We will give numerical estimates for
parameters of the microlensing effect. If the distance between a
dark body and the Sun is equal to $\sim 10$~kpc, then the angular
size of Einstein cone of the dark body with a solar mass is equal to
$\sim 0.001''$ or the linear size of Einstein cone is equal to about
10~AU. It is clear that since  angular distances between two images
are very small, it is very difficult  to resolve the images by
ground based telescopes at least in an optical band. It was a reason
that Einstein noted if gravitational lenses and sources are stars
and the separation angle between images is very small gravitational
lens phenomenon hardly ever be detectable
\cite{Einstein_36}.\footnote{Therefore, the microlensing effect is
observed analyzing a luminosity of a source as it was originally
proposed by Byalko \cite{Byalko69}.} However, recently, a direct
method to measure Einstein angle $\phi_E$ was proposed to resolve
double images generated by microlensing with an optical
interferometer (say VLTI) \cite{Delpl01}. Moreover, it was planned
to launch astrometrical space probes, such as US
SIM\footnote{http://sim.jpl.nasa.gov/whatis/.} and European
GAIA\footnote{http://astro.estec.esa.nl/GAIA.}, these instruments
will have accuracies about 10 microarcseconds and could resolve
image splitting for microlensing events. Applications of future
space missions for astrometrical microlensing searches are discussed
\cite{Zakharov_06}.
 An optical depth of microlensing for distant quasars
was discussed for different locations of microlenses
\cite{Zakharov04} (microlensing event candidates were found with 1.5
m RTT -150 telesope for gravitationally lensed system SBS 1520+530
\cite{Khamitov_06}). An influence of microlensing on spectral lines
and spectra in different bands was analyzed \cite{Popovic_ApJ_2006}.

Basic criteria for microlensing event identification are that a
light curve should be  symmetrical and achromatic. If we consider a
spherically symmetric gravitational field of a lens, a point source
and a short duration of microlensing event then the statement about
the symmetrical and achromatic light curves will be a correct claim,
but if we consider a more complicated distribution of a
gravitational lens field or an extensive light source then some
deviations of symmetric light curves may be observed and (or) the
microlensing effect may be chromatic \cite{Zakharov97}.

Many years  ago it was found that densities of visible matter is
about 10\% of total density in galactic halos  \cite{Oort32} and the
invisible component is called as dark matter (DM) and now it is
known that the matter density (in critical density units) is
$\Omega_m=0.3$ (including baryonic matter $\Omega_b \approx
0.05-0.04$, but luminous matter $\Omega_{\rm lum} \approx 0.001$),
$\Lambda$-term density $\Omega_\Lambda=0.7$ \cite{Spergel_06}. Thus,
baryonic density is a small fraction of total density of the
Universe. Probably galactic halos is "natural" places to store not
only baryonic DM, but non-baryonic DM as well. If DM forms objects
with masses in the range $[10^{-5},10]M_\odot$ microlensing could
help to detect such objects. Thus, before intensive microlensing
searches it was a dream that microlensing investigations could help
us to solve DM problem for Galactic halo at least.

As it was mentioned before, at the first time a possibility to
discover microlensing using observations of stellar light curves was
discussed in \cite{Byalko69} (however, to increase a probability in
the original paper it was proposed  to detect very faint flashes for
the background star light curves and in this form the idea hardly
ever is realizable). Systematic searches of dark matter using
typical variations of light curves of individual stars from millions
observable stars started after Paczynski's discussion of the halo
dark matter discovery using monitoring stars from Large Magellanic
Cloud (LMC) \cite{Paczynski86}.  In the beginning of the nineties
new computer and technical facilities providing the storage and
processing the huge volume of observational data were appeared and
it promoted at the rapid realization of Paczynski's proposal (the
situation was different in the Byalko's paper time). Griest
suggested to call the microlenses as Machos (Massive Astrophysical
Compact Halo Objects) \cite{Griest91}. Besides, MACHO is the name of
the project of observations of the US-English-Australian
collaboration which observed the LMC and Galactic bulge using 1.3 m
telescope of Mount Stromlo observatory in Australia.\footnote{MACHO
stopped since the end of 1999.} Since for the microlens searches one
can monitor several million stars for several years, the ongoing
searches have focused on two targets: a) stars in the Large and
Small Magellanic Clouds (LMC and SMC) which are the nearest galaxies
having lines of sight which go out of the Galactic plane and well
across the halo; b) stars in the Galactic bulge which allow to test
the distribution of lenses near to the Galactic plane (here paper we
do not discuss microlensing for distant quasars. The first papers
about the microlensing discovery were published by the MACHO
collaboration \cite{Alcock93} and the French collaboration EROS
(Exp\'erience de Recherche d'Objets Sombres)
\cite{Aubourg93}.\footnote {EROS experiment stopped in 2002
\cite{Moniez01}.}

First papers about the microlensing discovery toward Galactic bulge
were published by the US -- Polish Optical Gravitational Lens
Experiment (OGLE) collaboration, which used 1.3 m telescope at Las
Campanas Observatory. Since June 2001, after second major hardware
upgrade OGLE entered into its third phase, OGLE III as a result the
collaboration observes more than 200 millions stars observed
regularly once every 1 -- 3 nights. In last years OGLE III detected
more than four hundreds microlensing event candidates each year
\cite{Udalski2003}.\footnote{http://www.astrouw.edu.pl/~ogle/ogle3/ews/ews/html.}

MOA (Microlensing Observations in Astrophysics) is collaboration
involving astronomers from Japan and New Zealand
\cite{Bond01}.\footnote{http://www/roe.ac.uk/\%7Eiab/alert/alert/alert/html.}

To investigate Macho distribution in another direction one could use
searches toward M31 (Andromeda) Galaxy lying at 725~kpc (it is the
closest galaxy for an observer in the Northern hemisphere). On the
other hand, there are a number of suitable telescopes concentrated
in the Earth semisphere.
 In
 nineties two collaborations AGAPE (Andromeda Gravitational
 Amplification Pixel Experiment, Pic du Midi,
 France)\footnote{The POINT-AGAPE collaboration started in 1999 with the
  2.5 m Isaac Newton Telescope (INT)\cite{Kerins01_MNRAS},
 the new robotic project Angstrom
 was proposed as well \cite{Kerins06}.}
 and VATT
 started to monitor pixels instead of individual stars
 \cite{Moniez01,LeDu01}. These teams reported about discoveries of
 several microlensing event candidates \cite{Calchi_2005}.
Results of Monte Carlo simulations simulations for these
observations and differences between pixel and standard microlensing
are discussed \cite{Ingrosso_05}.

Concerning microlens detections one can say that even ten years ago
it was no doubt about this issue \cite{Paczynski96}. However, it is
impossible to say exactly which part of the microlensing event
candidates is actually connected with the effect since probably
there are some variable stars among the event candidates, it could
be stellar variability of an unknown kind.\footnote{ The
microlensing event candidates proposed early by the EROS
collaboration ( \#1 and \#2) and by the MACHO collaboration (\#2 and
\#3) are considered now as the evidence of a stellar variability
\cite{Paczynski96}.} Below we will list the most important results.
Observed light curves are  achromatic and their shapes are
interpreted by simple theoretical expressions very well, however,
there is not complete consent about "very well interpretation" since
even for the event candidate MACHO \# 1 the authors of the discovery
proposed two fits. Dominik and Hirshfeld  suggested that the event
could be fitted perfectly in the framework of the binary lens model
\cite{Dominik94}, but one can assume that the microlensing event
candidate could be caused by a non-compact microlens
\cite{Gurevich96}.

Using photometric observations of the caustic-crossing binary lens
microlensing event EROS BLG-2000-5, PLANET collaboration reported
about the first microlens mass determination, namely the masses of
these components are 0.35~$M_\odot$ and 0.262~$M_\odot$ and the lens
lies within 2.6~kpc of the Sun \cite{An02}.

Gravitational microlensing events due to stellar mass black holes
have been discovered \cite{Bennett02}. The lenses for events
MACHO-96-BLG-5 and MACHO-96-BLG-6 are the most massive, with mass
estimates $M/M_\odot=6^{+10}_{-3}$ and $M/M_\odot=6^{+7}_{-3}$,
respectively, however later it was established that event
MACHO-99-BLG-22 is a strong BH candidate (78\%), MACHO-96-BLG-5 is
marginal BH candidate (37 \%), but MACHO-96-BLG-6 is a weak BH
candidate (2\%) \cite{Poindexter_05}.



The optical depth towards the Galactic bulge is equal to $
\sim 3 \times 10^{-6} $, so it is larger than the estimated value
\cite{Alcock00a}.


Analysis of 5.7 years of photometry on 11.9 million stars in LMC by
MACHO collaboration reveals 13 -- 17 microlensing events
\cite{Alcock00b}. The optical depth towards the LMC is equal to
$\tau(2 < \hat{t} < 400 {\rm~ days}) =1.2^{+0.4}_{-0.3} \times
10^{-7} $, so, it is smaller than the estimated  value. The maximum
likelihood analysis gives a Macho halo fraction $f=0.2$. Estimates
of the following probabilities $P (0.08 < f <0.5)=0.95$ and $P(f=1)
< 0.05$ are given. The most likely Macho mass $M \in [0.15, 0.9]
M_\odot$, depending on the halo model and total mass in Machos out
50~kpc is found to be $9^{+4}_{-3} \times 10^{10} M_\odot$. EROS
collaboration gives a consistent conclusion, namely, this group
estimates the following probability $P (M \in [10^{-7},
1]M_\odot~\&~f>0.4) <~ 0.05$ \cite{Lasserre00}. Recently the
collaboration concluded that the optical depth toward LMC is $\tau <
0.36 \times 10^{-7}$ (95\% C.L.) it means that macho contribution in
halo mass is less than 7 \% \cite{Tisserand_06}. However, these
conclusions are based on assumptions about mass and spacial
distributions of microlenses and these distributions are not known
very well and in principle microlensing searches is realistic way to
improve the knowledge, but in this case we need thousands of events.

Since an existence of planets leads to a formation of fold and cusp
type caustics \cite{Zakharov_95}, one can detect extra peaks due to
caustic crossing by a background star as a result of a proper
motion.  Among other important discoveries one should point out
planet detections done the with the microlensing technique
\cite{Abe_2004} such the discovery of planet with 5.5 Earth masses
because that is the lightest extrasolar planet discovered at the
moment taking into account all the techniques used for extrasolar
planet searches and it means that existence of cool rocky planets is
common phenomenon in the Universe \cite{Beaulieu_06}. Very recently
the Neptune like planet was discovered by the same manner
\cite{Gould_06}.

When new observational data would be collected and the processing
methods would be perfected, probably some microlensing event
candidates lost their status, but perhaps new microlensing event
candidates would be extracted among analyzed observational data.
Thus, the general conclusion may be done, the very important
astronomical phenomenon was discovered, but some quantitative
parameters of microlensing will be specified in future. However, the
problem about  a content of 80\% (or even 93\% according to EROS
point of view)¡¡of DM in the halo of our Galaxy is still open
(before microlensing search are people hoped that microlensing could
give an answer for this problem). Thus, describing the present
status Kerins wrote adequately that now we have "Machos and clouds
of uncertainty" \cite{Kerins01}. It means that there is a wide field
for studies, in particular, pixel microlensing, microlensing for
gravitational lensed systems and extrasolar planet searches seem to
be the most promising issues.


The author is indebted to D.~Blaschke for his kind invitation to
present a lecture on this subject at the Helmholtz International
Summer School "Dense Matter In Heavy Ion Collisions and
Astrophysics", J.~Wang and J.~Zhang for the hospitality in NAOC and
important discussions. AFZ is also grateful to the National Natural
Science  Foundation of China (NNSFC) (Grant \# 10233050) and the
National Basic Research Program of China (2006CB806300) for a
partial financial support of the work.

\end{document}